\pdfoutput=1
\documentclass[%
reprint,
amsmath,amssymb,  aps, prc, nofootinbib
]{revtex4-1}
\usepackage{mathtools}
\usepackage{graphicx}
\usepackage{dcolumn}
\usepackage{bm}
\usepackage{subfigure} 
\usepackage{xcolor}

\newcommand{\mref}[1]{\textbf{\textsc{\textcolor{red}{[REF]}}}}

\begin{document}

	\title{Global angular momentum generation in heavy-ion reactions within a hadronic transport approach}
	\author{Nils Sass$^{1}$}  
	\author{Marco M\"uller$^{1} $}
	\author{Oscar Garcia-Montero$^{1,2}$} 
	\author{Hannah Elfner$^{3,1,4,5}$}

	\affiliation{$^1$Institute for Theoretical Physics, Goethe University,
		Max-von-Laue-Strasse 1, 60438 Frankfurt am Main, Germany}
	\affiliation{$^2$Fakult\"at f\"ur Physik, Universit\"at Bielefeld, 33615 Bielefeld, Germany}
	\affiliation{$^3$GSI Helmholtzzentrum f\"ur Schwerionenforschung, Planckstr. 1, 64291
		Darmstadt, Germany}
	\affiliation{$^4$Frankfurt Institute for Advanced Studies, Ruth-Moufang-Strasse 1, 60438
		Frankfurt am Main, Germany}
	\affiliation{$^5$Helmholtz Research Academy Hesse for FAIR (HFHF), GSI Helmholtz Center,
		Campus Frankfurt, Max-von-Laue-Straße 12, 60438 Frankfurt am Main, Germany}
	
	\date{\today}
	
	\begin{abstract}
	 In 2017, the STAR collaboration at the Relativistic Heavy Ion Collider (RHIC) has measured finite global angular momentum in heavy-ion collisions through a spin polarization measurement of $\Lambda$ hyperons. This measurement revealed a high angular momentum of the heavy ions and provided experimental evidence for vorticity in the quark-gluon plasma (QGP) for the first time. In order to investigate the underlying mechanisms, a dynamic description of the transfer of angular momentum is required. In this work, the microscopic non-equilibrium transport approach SMASH (Simulating Many Accelerated Strongly-interacting Hadrons) is applied to study the generation of global angular momentum by the interaction of two nuclei. As SMASH provides access to the whole phase-space evolution of every particle at any given time, it allows to assess the fraction of angular momentum generated in the fireball by all participants. We confirm the previous modeling by Becattini \textit{et al} within a geometric Glauber model approach, which found that the angular momentum transfer reaches a unique maximum in mid-central collisions during time evolution. The corresponding impact parameter is around $b=4-6$ fm for all beam energies from $\sqrt{s_{\rm NN}}=2.41-200$ GeV. 
	 Even though angular momentum is not conserved locally in the transport approach a priori, we identify the contributions to the conservation violation and propose optimal setups for different energy regimes that recover conservation, based upon the test particle method and the treatment of Fermi motion. Furthermore, the system size and centrality dependence are investigated. 
\end{abstract}

	\maketitle
	
	\section{Introduction}
	\label{intro}
	
	The study of relativistic collisions of heavy ions at various beam energies provides a unique tool to study the fundamental properties of strongly interacting matter in different regions of the QCD phase diagram \cite{Busza:2018rrf}. Some of these regions may be accessed with current and future facilities dedicated to the experimental realization of heavy-ion collisions (HICs). At low to intermediate beam energies these include the HADES experiment at the Facility for Antiproton and Ion Research (FAIR) \cite{Wilczek:2010ae} or the NICA experiment at the Joint Institute for Nuclear Research (JINR) \cite{article}, as well as the Beam Energy Scan program by the Relativistic Heavy Ion Collider (RHIC). The low energy regime is well known to be dominated by hadronic interactions and, therefore, it is successfully described by transport approaches that consider hadronic degrees of freedom \cite{Weil:2016zrk, Buss:2011mx, EHEHALT1996449, Nara:1999dz, Bass:1998ca}.
	
	In this range of collisional energies, the experimental detection of a non-vanishing polarization of $\Lambda$ hyperons lead to the breakthrough discovery that the medium created during heavy-ion collisions has a strong vortical structure \cite{Karpenko:2017lyj,Jiang:2016woz,Becattini:2015ska,McInnes:2018ibt,HADES:2022enx}. This polarization is induced by the spin-orbit coupling of the hyperons to the non-zero local rotation, the so called vorticity \cite{Liang:PolarizedQGP,Gao:2007bc,Betz:Polarization}.
	This can be directly thought to be a consequence of the initial configuration of the ions, which travel close to the speed of light generally with a non-vanishing offset distance, the so-called impact parameter. This initial condition grants the system a large amount of initial orbital angular momentum. For example, an AuAu collision at $\sqrt{s_{\rm NN}} = 39 \text{ GeV}$ and an impact parameter of $b=5$ fm carries an initial angular momentum of $|L_0| \sim A\,b\sqrt{s_{\rm NN}}/2\sim 2 \times 10^5\hbar$, where $A$ denotes the atomic mass number.
	As the early out-of-equilibrium medium created by the collision evolves, such large global angular momentum is rapidly broken down into smaller rotational domains, which can eventually be described in the hydrodynamical limit by vorticity. 
	
	While the total angular momentum is fully conserved and it scales almost linearly with the energy of the collision, its deposition in the interacting region exhibits a more complex behavior. As the energy increases, the relative deposition of angular momentum into the medium is reduced, making the nuclei more transparent \cite{Becattini_2008}. There is therefore a maximum angular momentum deposition to energy ratio, which can be traced to be within the low-to-intermediate energy regime. This argument is supported by the energy dependence of the measurements of the $\Lambda$ hyperon polarization, where the total polarization signal decreases with the collision energy \cite{STAR:2017ckg}. Understanding the dynamics of angular momentum at intermediate beam energies is not only important to guide experimental programs in selecting systems and centralities for the deeper understanding of observables related to angular momentum, but also provides constraints on the dynamical evolution itself. Well understood dynamical approaches are the basis for the future solid identification of phase transition signals. 
	
	In this work we investigate the dynamical deposition of angular momentum into the medium using the hadronic transport approach SMASH \cite{Weil:2016zrk,SMASHweb}. This grants us the possibility to access the out-of-equilibrium nature of the deposition of energy and angular momentum in the initial stages of a heavy-ion collision. Even though purely hadronic dynamics is restricted to low beam energies, we can use a hadron-string approach at higher beam energies. This renders the full approach valid for a large range of collisional energies, opening the window to study the energy dependence of angular momentum. Additionally, this kinetic approach allows us to build on previous geometric studies \cite{Becattini_2008} by not only being able to assess the conversion of initial angular momentum to the medium, but also to resolve the evolution of angular momentum with time.
	
	This work is organized as follows: In section \ref{sec:model} we briefly review the hadronic transport model SMASH, the basics of angular momentum transfer, as well as giving the definitions we use to quantify the in-medium angular momentum. In section \ref{sec:results} we present the results of the study, where we explore and discuss the impact parameter, energy and system size dependence of the deposited angular momentum. We also discuss the evolution of angular momentum, total and in-medium. As it is widely known \cite{sym13101852,Cheng_2002,Kodama:Causality,Zhang:Causality}, angular momentum conservation is broken in transport approaches. However, we not only show that the local broken angular momentum conservation has only a minor impact on the overall evolution of angular momentum, but also that our results display a good qualitative agreement with theoretical predictions and experimental data. Finally we present a short summary and outlook.

	\section{Angular momentum in SMASH}
	\label{sec:model}
	
	In this work, the model applied to study angular momentum transfer in heavy-ion collisions is the hadronic transport approach SMASH in version 2.0 \cite{Weil:2016zrk, SMASHweb}, which has been shown to be successful in the description of particle production in numerous setups \cite{Weil:2016zrk,Staudenmaier:2020xqr,Mohs:2019iee,Steinberg:2019wgm}. It provides the full phase-space evolution of every particle by providing an effective solution of the relativistic Boltzmann equation
	\begin{equation}\label{Boltzmann Equation}
		p^\mu \partial_\mu f_i(x,p)+m_i F^\alpha\partial^p_\alpha f_i(x,p) = C^i_\text{coll}
	\end{equation}
	in the non-equilibrium dynamics regime, where $C^i_\text{coll}$ denotes the collision term, $F^\alpha$ is an external force acting on each particle species $i$ with mass $m_i$ and its corresponding single particle distribution $f_i(x,p)$.
	The relevant degrees of freedom in SMASH include most of the well-established hadrons and hadronic resonances listed in the Particle Data Group \cite{Zyla:2020zbs}, up to a mass of $m\approx 2.3\,\text{GeV}$. In SMASH, resonances are treated with their vacuum spectral functions according to a Breit-Wigner distribution with mass-dependent widths following the Manley-Saleski ansatz \cite{Manley:1992}. The interface to high-energy hadronic interactions, by means of hard scattering processes and string fragmentation, is realized within the string model Pythia 8 \cite{Sjostrand:2006za,Sjostrand:2007gs}. A detailed overview of our approach including a complete list of incorporated degrees of freedom can be found in Ref. \cite{Weil:2016zrk}. 
	
	At the basis of each interaction is the collision search based on the total cross-sections. The geometric collision criterion, as it is employed in the UrQMD (Ultra-relativistic Quantum Molecular Dynamics) approach \cite{Bass:1998ca,Bleicher:1999xi} is based upon determining the transverse distance of closest approach $d_T$ between two incoming particles. If $d_T$ undershoots a threshold distance $d_{int}$,
	\begin{equation}\label{collision criterion}
		d_T<d_{int}=\sqrt{\frac{\sigma}{\pi}}\, ,
	\end{equation}
	given by a geometrical interpretation of the corresponding total cross-sections, the particles collide. In this work we apply a modified version of the geometric collision criterion in a covariant formulation, as presented in Ref. \cite{Hirano:2012yy}.
	Eq. (\ref{collision criterion}) is responsible for the well-known broken angular momentum conservation in transport approaches, as it encodes an immediate finite range interaction which breaks Poincaré invariance \cite{sym13101852,Cheng_2002,Kodama:Causality,Zhang:Causality}. As a direct consequence, angular momentum cannot be conserved locally. However, the concept of test particles provides us with a tool to restore locality and remedy angular momentum conservation in the limit $N_{\rm test}\rightarrow \infty$.
	
	SMASH applies the test particle method, which means that cross-sections $\sigma$ and initial particle number $N$ are scaled as
	\begin{align}\label{test_particle_method}
		N &\rightarrow N_{\text{test}} N\\
		\sigma &\rightarrow N^{-1}_{\text{test}} \sigma \label{testparticle_sigma_scaling}
	\end{align}
	while keeping the scattering rate fixed. The factor $N_{\text{test}} $ thereby defines the number of test particles. Increasing N has two effects in particular. On the one hand, the cross-sections are decreased, which causes interactions to become more local and suppresses non-conservation of angular momentum. The corresponding increase of particle number results in a significant increase in computational time. For the following considerations, a test particle number of $N_{\text{test}}=20$ is chosen as a compromise between relevance for the results and computational time.
	
	Furthermore, to simulate nucleus-nucleus collisions, it is necessary to define initial conditions for the phase-space distributions of the initial nuclei. In coordinate space, spherical nuclei (e.g. gold) are sampled with nucleons according to a Woods-Saxon profile \cite{Woods:1954}.\\
	The momentum space distribution of the nuclear ground state, on the other hand, gives rise of the so-called Fermi motion. This distribution is reflected in a uniformly filled sphere in momentum space, also known as Fermi sphere, with radius
	\begin{equation}\label{Fermi Radius}
		p_F(\textbf{r})=\hbar\,c\left(3\pi^2 \rho(\textbf{r}) \right)^{1/3}\,,
	\end{equation}
	where $\rho(\textbf{r})$ is the spatial nucleon density at the point $\textbf{r}$. The Fermi momentum will typically be of order $p_F\approx 300\,\text{MeV}$ which corresponds to an excess kinetic energy of roughly $p_F^2/2m_N \sim 45\, \text{MeV}$. Even though the energy contribution originating from Fermi momenta is comparably small, for low beam energy setups it is important to take it into account as it gives a non-negligible contribution to the transverse mass spectra. In this work, we apply the frozen Fermi approximation in which Fermi momenta are neglected during propagation and the additional momentum is only considered for the collisions. The Fermi motion contributes to the total angular momentum of the system as will be shown in the following.\\
	
	To study the evolution of angular momentum in heavy-ion collisions, the relevant quantities are the initial angular momentum $L_0$, defined as the total angular momentum $L_{\text{tot}}$ of all nucleons at time $t<0\,\text{fm}$, and the final angular momentum $L_f$ at a time when no secondary collisions occur any more. Unless stated otherwise, we choose the final time to be $t_F=200\,\text{fm}$. To investigate which fraction of the initial angular momentum is transferred to the fireball, we further distinguish between the remaining angular momentum $L_r$ in the interaction medium carried by the participants and the contribution of spectators $L_{\text{sp}}$, such that 
	\begin{equation}
		L_{\text{tot}}=L_{\text{r}}+L_{\text{sp}}\,.
	\end{equation}
	The generation of angular momentum in a heavy-ion collision is predominantly driven by the initial geometry of the collision setup which is dictated by the chosen impact parameter. In SMASH, nuclei are propagated along the z-axis while a non-zero impact parameter creates a relative offset between the nuclei centers in x-direction. Consequently, the main contribution of the orbital angular momentum points in the y-direction of the computational frame and all other components are negligible for qualitative analysis. 
	\begin{figure}[t]
		\hspace{-10pt}
		\includegraphics[width=8.9cm]{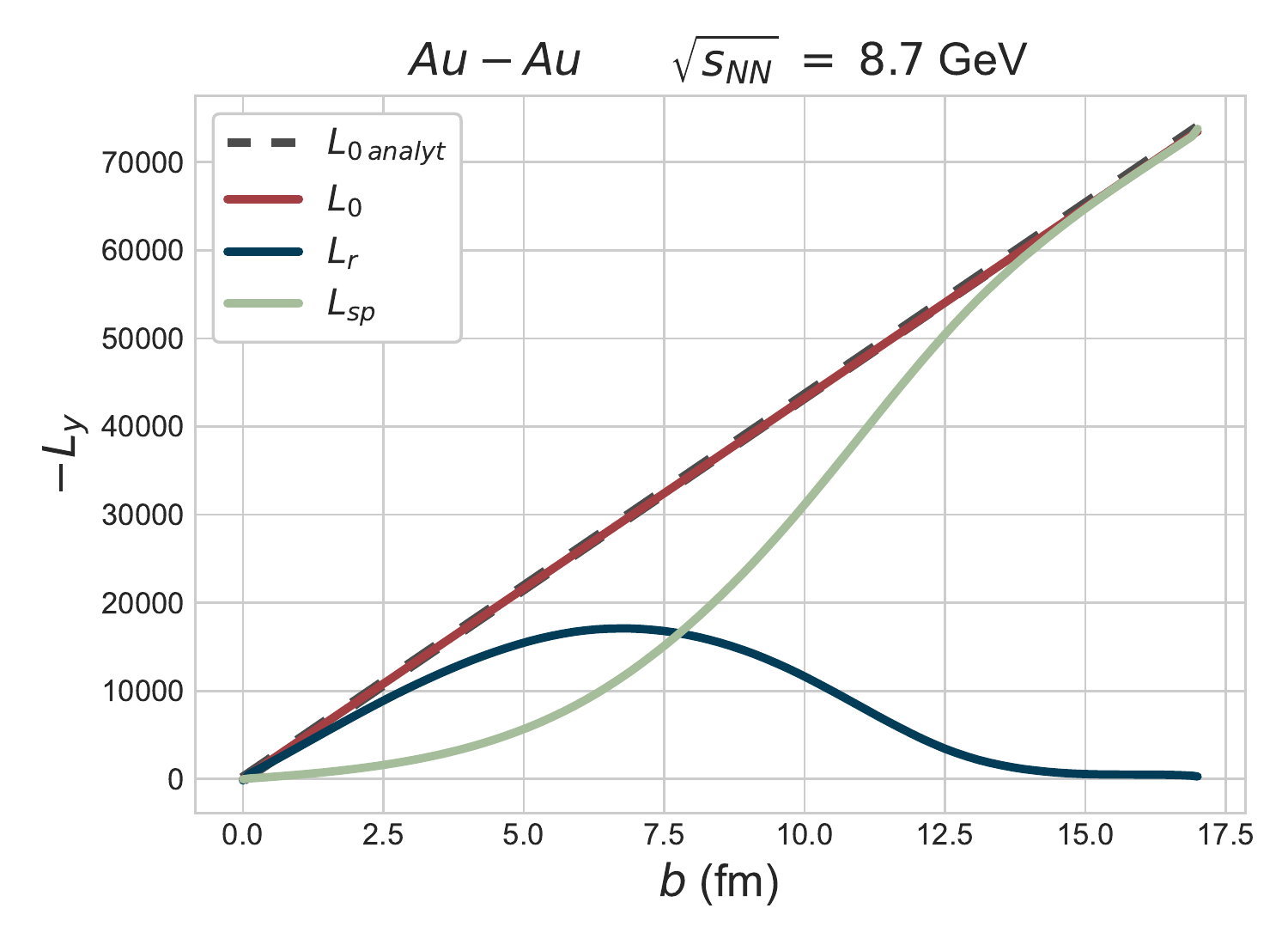}
		\caption{Angular momentum as a function of impact parameter for AuAu collisions at $\sqrt{s_{\rm NN}}= 8.7$ GeV. Shown is the angular momentum of the participants, spectators, the total value and the analytical approximation of the total angular momentum.}
		\label{fig:total_spec87}
	\end{figure}
	\begin{figure}
		\centering
		\includegraphics[width=8.6cm]{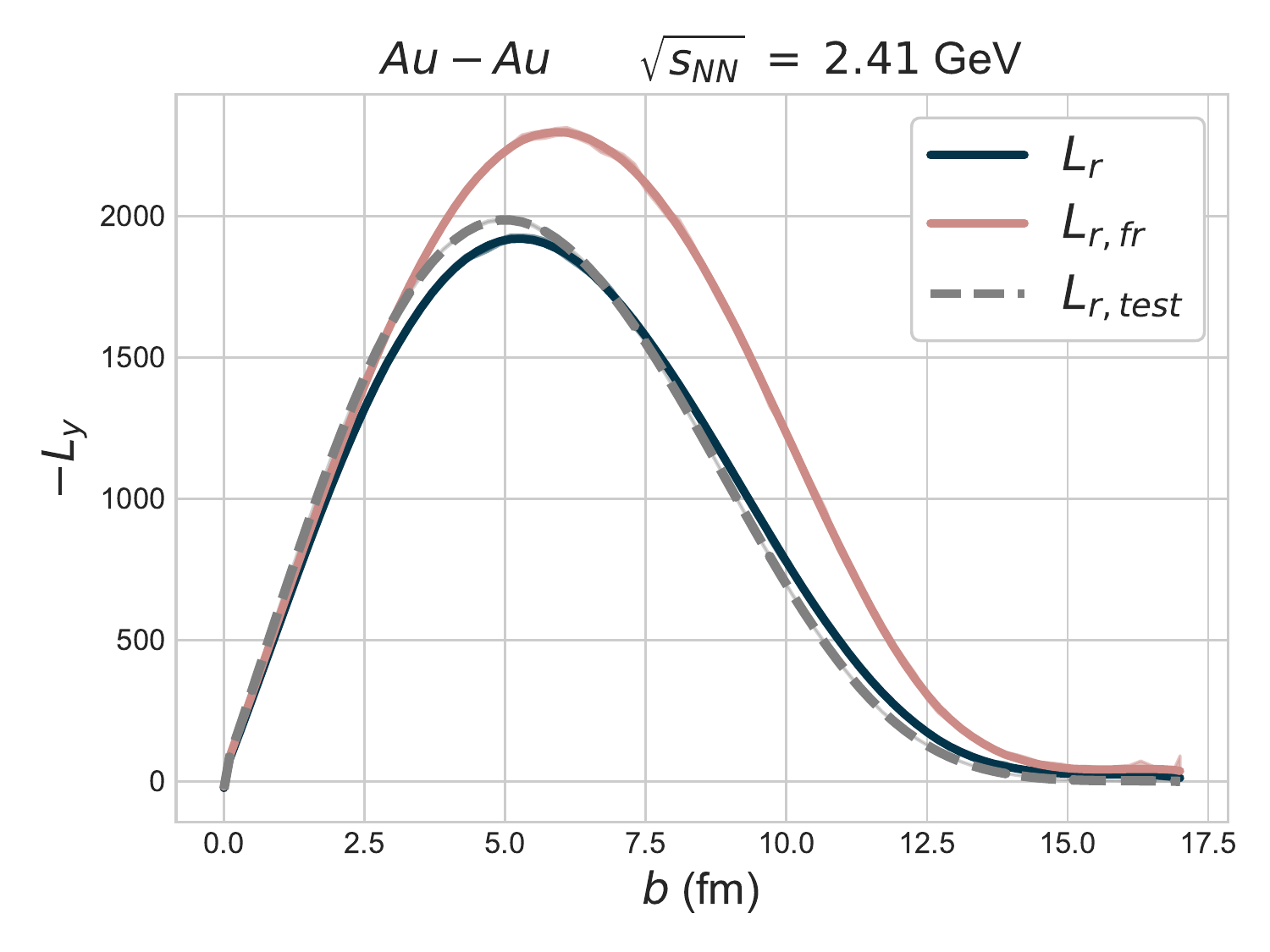}\vspace{28pt}
		\includegraphics[width=8.6cm]{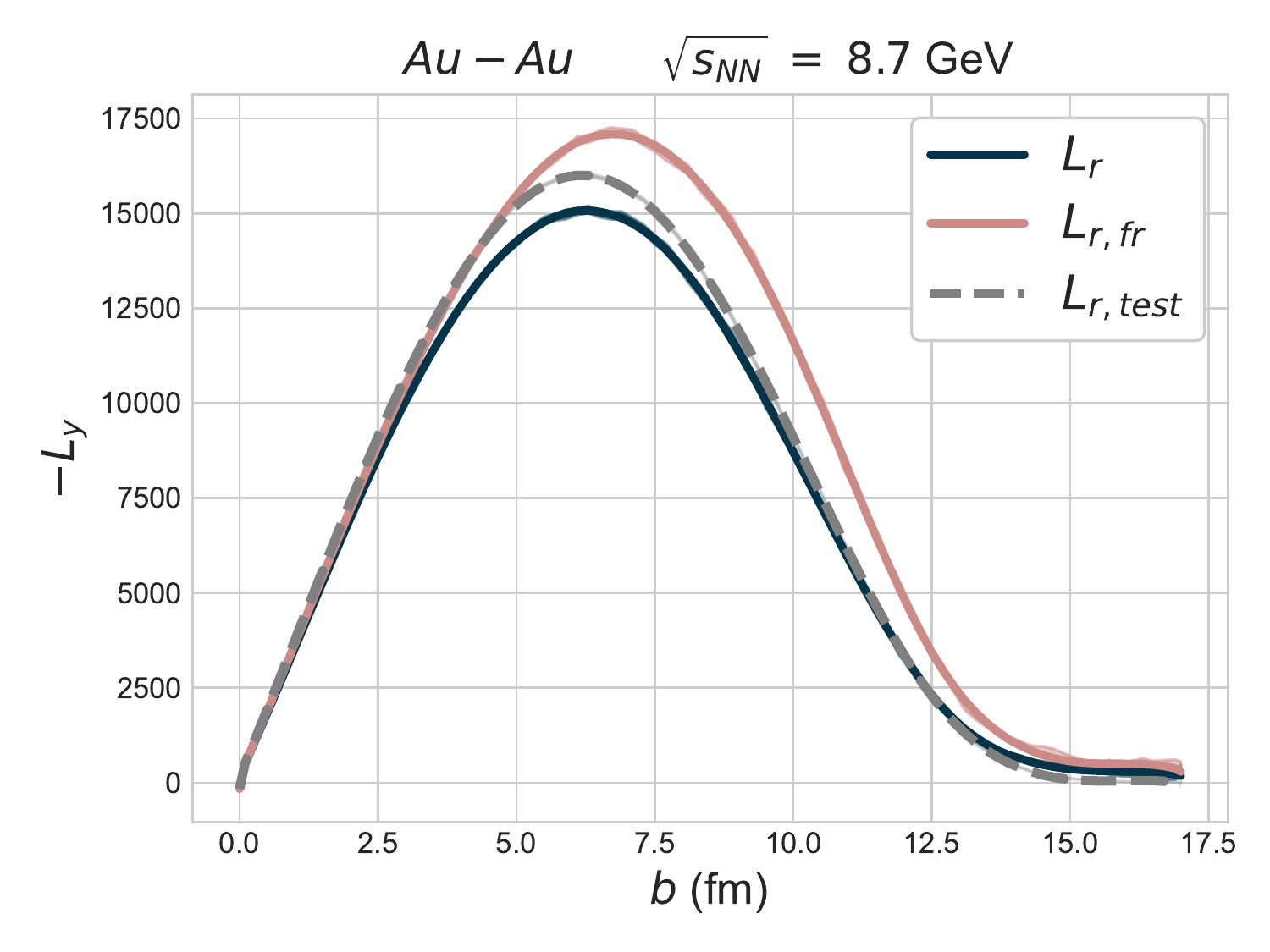}\vspace{27pt}
		\includegraphics[width=8.6cm]{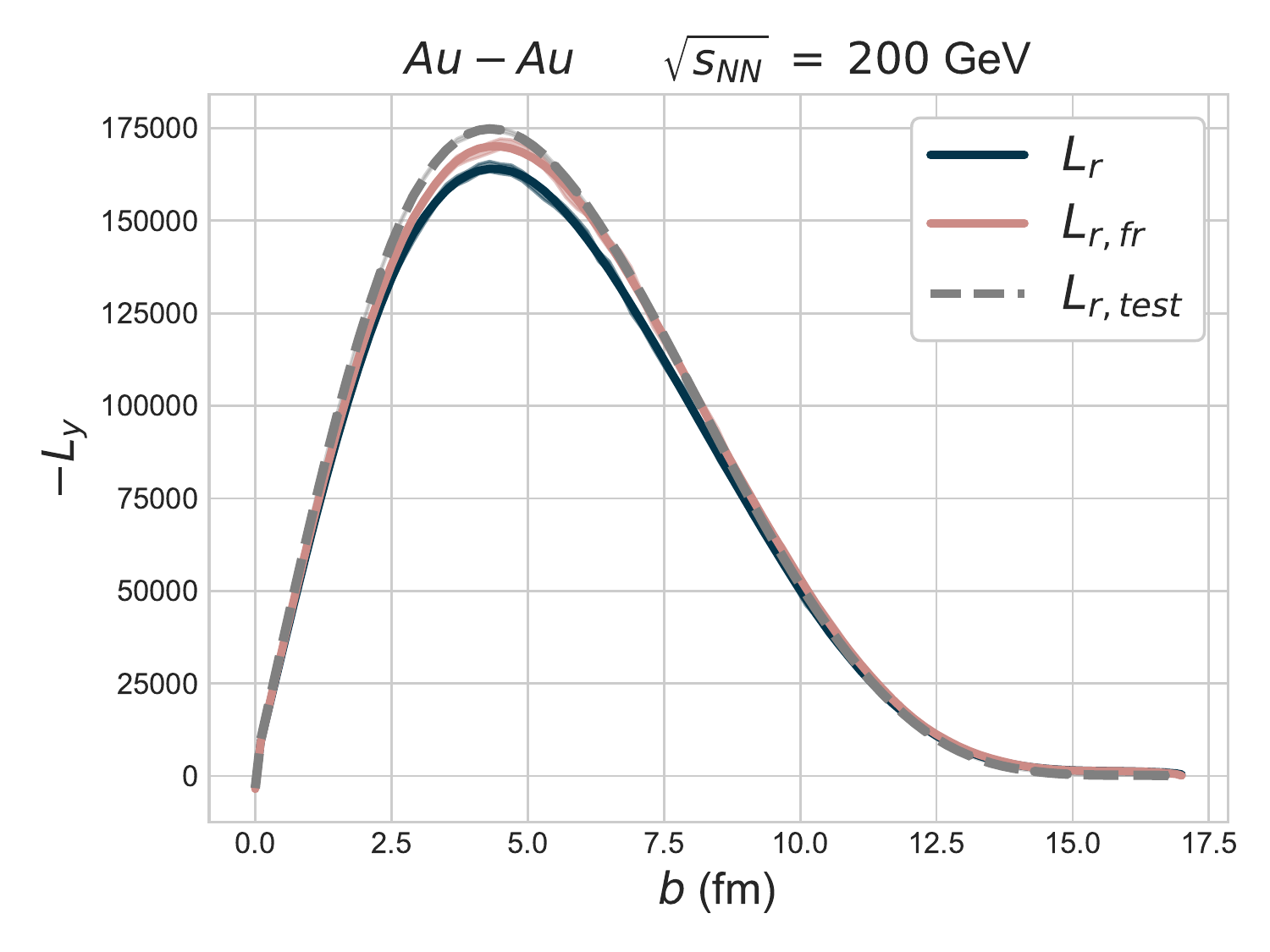}
		\caption{Angular momentum as a function of impact parameter for AuAu collisions at $\sqrt{s_{\rm NN}}=2.41, 8.7$ and $200$ GeV within the fireball. Compared are three calculations with and without Fermi motion, as well as including 20 test particles depicted by the gray dashed line.}
		\label{fig:imp_energy}
	\end{figure}
	We can estimate the behavior of the orbital angular momentum $L_y$ as function of the invariant mass $\sqrt{s}$ by considering two identical initial nuclei with 4-momenta $P=(\sqrt{s}/2, 0, 0, m \gamma v_z)^T$. The square of this 4-momentum then determines the z-component of the velocity and the $\gamma$-factor
	\begin{equation}
		v_z^2=1-\frac{4m^2}{s}\,, \qquad \gamma=\frac{\sqrt{s}}{2m}\,,
	\end{equation}
	where we used that for two identical nuclei $\sqrt{s}=2\gamma m$. For large energies, i.e. $s >> 4m^2$, the orbital angular momentum becomes a linear function in $\sqrt{s}$:
	\begin{equation}\label{L as function of sqrt(s)}
		L_y=-\frac{b \sqrt{s}}{2} \sqrt{1-\frac{4 m^2}{s}} \approx -\frac{b}{2} \sqrt{s}\,.
	\end{equation}
	This result provides us with an estimate against which we can test our results from the next section for plausibility.
	\section{Results for global angular momentum} \label{sec:results}
	Unless otherwise stated, the following results obey the convention that all plots show as default our findings for which the Fermi motion was turned off.
	
	\subsection{Impact parameter dependence}\label{subsec:results_impact_parameter}
	In this section, we study the orbital angular momentum of AuAu collisions for different centralities and energies, ranging from $\sqrt{s_{\rm NN}}=2.41\,\text{GeV}$ to $\sqrt{s_{\rm NN}}=200\,\text{GeV}$.
	
	In Fig. \ref{fig:total_spec87} we show our results for the angular momentum of an AuAu collision at $\sqrt{s_{\rm NN}}=8.7\, \text{GeV}$ as a function of the impact parameter including frozen Fermi motion. 
	The blue and green curves show the angular momentum carried by participants and spectators, respectively. The total angular momentum, depicted by the red line, on the other hand, is compared to the dashed gray line, which shows the analytical estimate from Eq. (\ref{L as function of sqrt(s)}).
	Our findings for the fraction of angular momentum contained in the interaction region show a distinct maximum in mid-central collisions $b_{max}\approx 6.7\, \text{fm}$, while decreasing for even larger impact parameters until the angular momentum generation drops to zero beyond $b=17.0\text{ fm}$, where the nuclei no longer have an overlap on average. This result coincides with findings in \cite{Karpenko:2016jyx} and with predictions by Becattini \textit{et al} \cite{Becattini_2008,Becattini_2020} that have shown the same qualitative behavior in a geometric Glauber model approach. The maximum of $L_r$ thereby indicates the configuration for which the orbital angular momentum in the impact area is largest and for which we therefore expect the strongest vorticity in the QGP \cite{Becattini_2008, Deng_2016,STAR:2017ckg}. Considering also spectators, the total angular momentum of the whole system, as expected, follows the linear behavior we predicted with Eq. (\ref{L as function of sqrt(s)}). In general, the global angular momentum has to be zero in central collisions due to symmetry arguments. It is then increasing for mid-central collisions due to the off-set of the nuclei that work like spinning a top, while going to peripheral collisions the interaction rate decreases and eventually no angular momentum is transferred anymore to the fireball. 
	
	To understand the impact of Fermi motion on the remaining angular momentum in the medium, we compare calculations for AuAu collisions with and without frozen Fermi motion for different beam energies in Fig. \ref{fig:imp_energy}. Displayed as curves are fits to our calculations of the impact parameter dependence of the angular momentum of the participants. The blue curve shows the angular momentum of the system as function of the impact parameter when Fermi motion is turned off, while the dashed gray curve additionally includes $N_{\rm test}=20$ test particles. With the pink curve we show our results without test particles but taking Fermi motion into account. In general, the curves qualitatively show the same characteristic shape for all energies, i.e. for every beam energy we find a unique impact parameter $b_{max}$ for which $L_r$ reaches its maximum. 
	Notably, the inclusion of Fermi motion bestows a contribution to the orbital angular momentum. Despite the isotropic sampling of Fermi momenta in the nucleus rest frame, this isotropy is disrupted upon transitioning to the calculation frame due to a Lorentz boost along the z-direction. This boost sustains isotropy within the transverse plane while breaking it along the z-axis, leading to a discernible non-zero contribution to $-L_y$ within the interaction region.
	The pink curve lies above the blue curve for all values of $b$ and for all calculated beam energies, while the relative difference between both curves becomes smaller for higher beam energies as expected. Besides increasing the absolute value of angular momentum, Fermi motion also slightly shifts $b_{max}$ towards more peripheral collisions. Since in the frozen Fermi approximation Fermi momenta are considered in the moment of collision, they increase the momentum diffusion at the edge of the interaction region turning some spectators into participants and thus effectively increase the overlap area. In turn, the maximum impact parameter is shifted to slightly larger values.
	
	In contrast, with test particles that are incorporated to ensure conservation of angular momentum locally, we see that $b_{max}$ is pushed towards more central collisions. Following Eq. (\ref{testparticle_sigma_scaling}), the cross-sections inversely scale with $N_{\rm test}$ making interactions more local. The reduced interaction range has to be balanced by moving the nuclei closer to each other to recover the maximum angular momentum transfer. Overall the qualitative behavior and even the magnitude of angular momentum is only mildly affected at all beam energies. 
	
	\subsection{Time evolution} \label{subsec:time_evolution}
	\begin{figure}
		\centering
		\includegraphics[width=8.6cm]{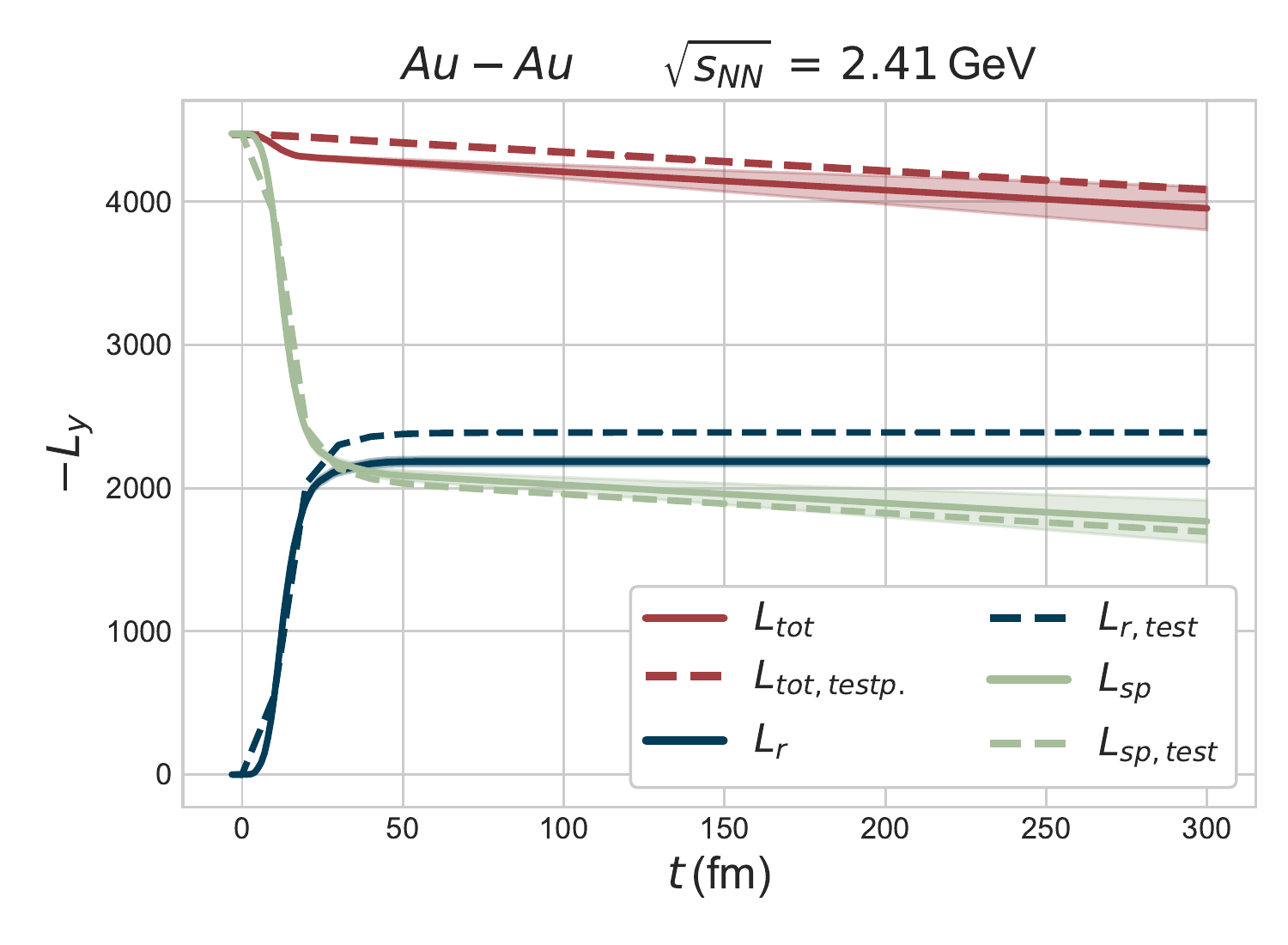}\vspace{24pt}
		\includegraphics[width=8.6cm]{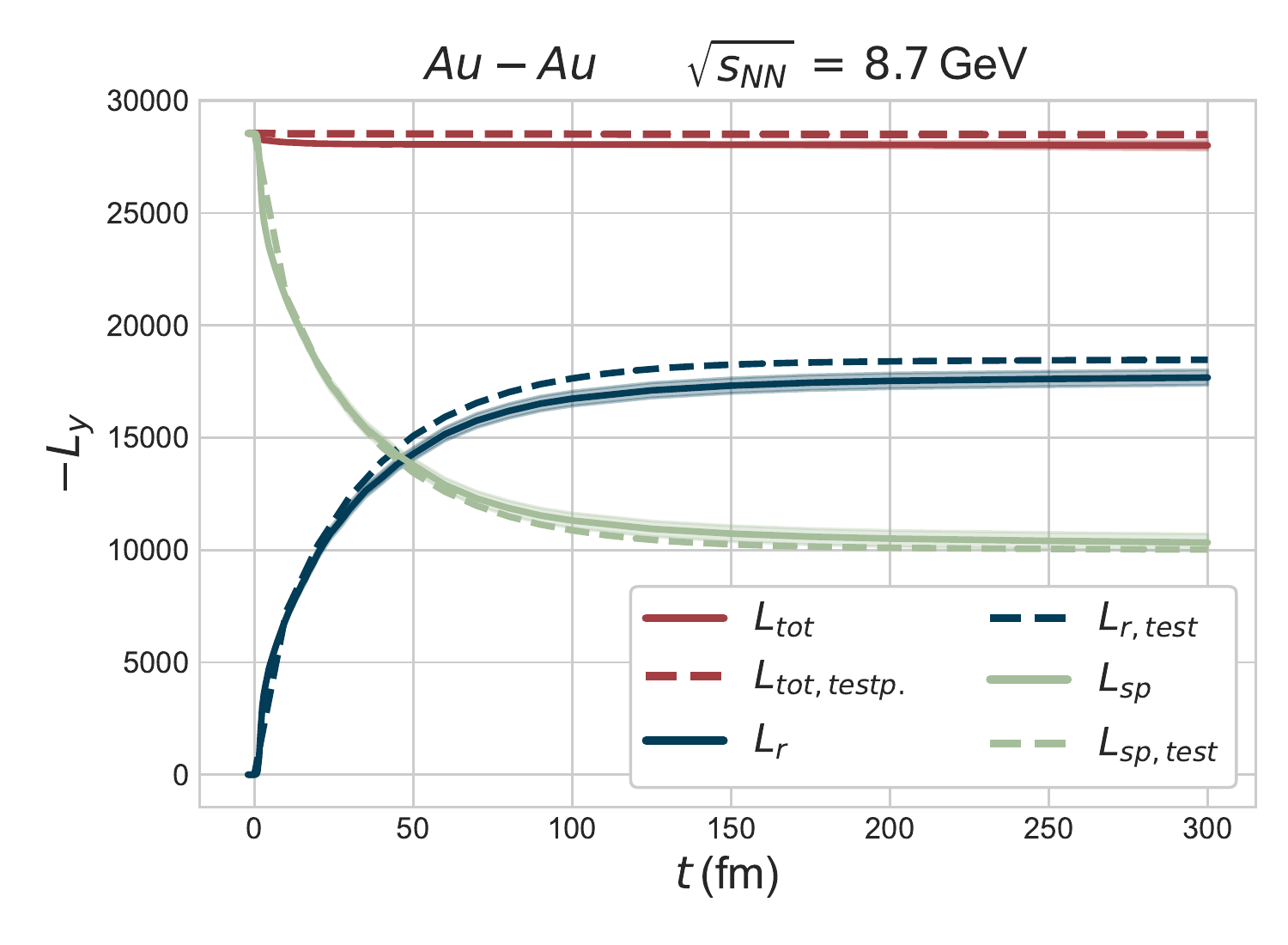}\vspace{24pt}
		\includegraphics[width=8.6cm]{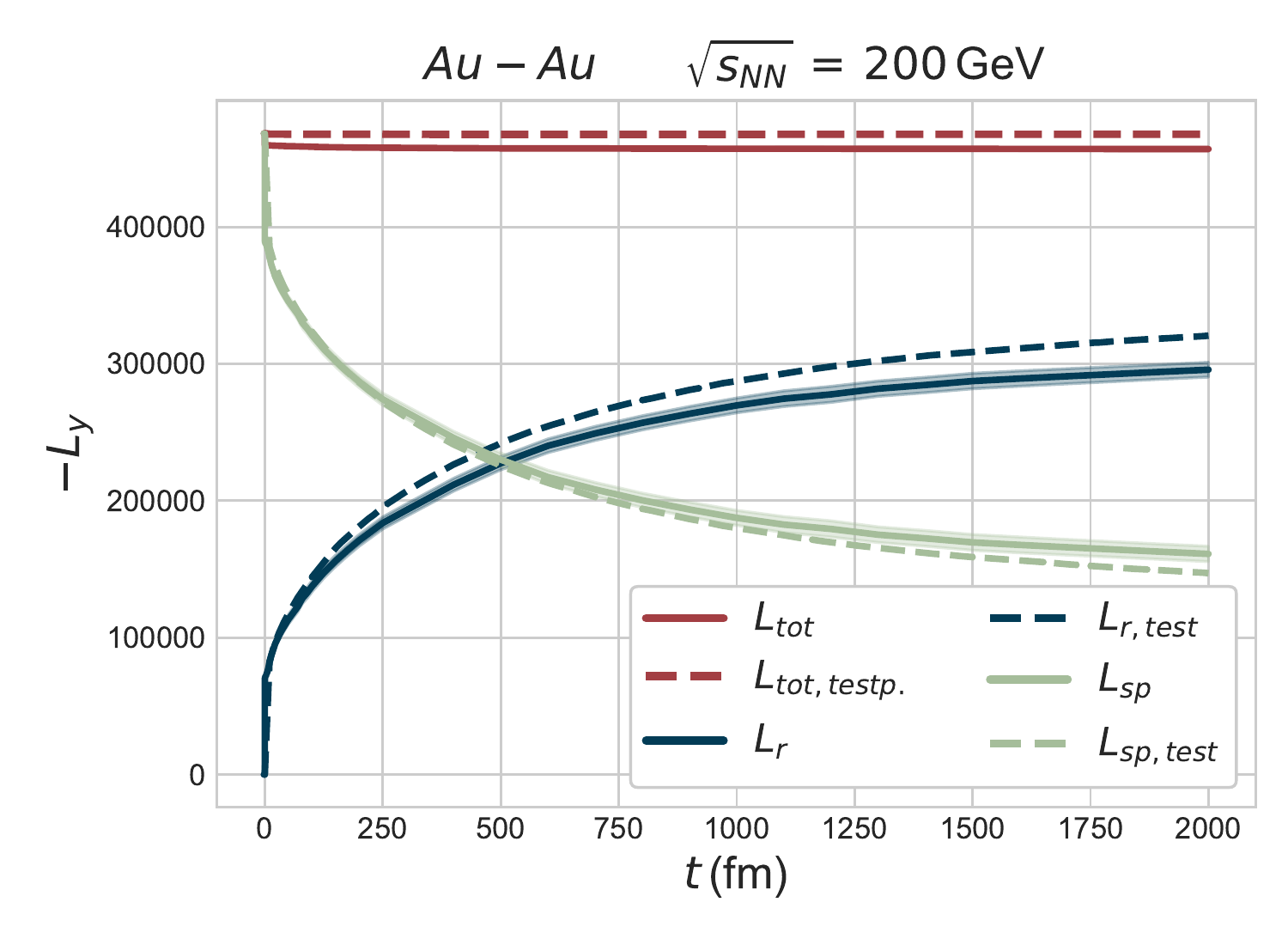}
		\caption{Time evolution of the angular momentum in AuAu collisions at $\sqrt{s_{\rm NN}}=2.4, 8.7$ and $200$ GeV including frozen Fermi motion. For each energy, the corresponding impact parameter was chosen such that it maximizes the angular momentum deposition in the interaction region. The dashed lines depict the result for the same simulations including 20 test particles.}
		\label{fig:time_evolution}
	\end{figure}
	With each interaction of two or more nucleons, the orbital angular momentum of the initial state is dynamically transferred to the participants of the heavy-ion collision over time until all secondary collisions have ceased. 
	In particular, this means that we expect a large deposition of angular momentum at times of highest interaction rates, which can be traced to the initial to very early collision stages for which a transport description is still valid. Although SMASH does not incorporate QGP formation, it is still relevant to understand how angular momentum is generated in initial hadronic interactions.\\
	Fig. \ref{fig:time_evolution} shows the time evolution of angular momentum in an AuAu collision with frozen Fermi motion at the impact parameter where the highest angular momentum transfer happens within the hadronic transport approach.
	As in Sec. \refeq{subsec:results_impact_parameter}, the red curve indicates the total angular momentum of the system, while the blue and green curves show the angular momentum evolution of participants and spectators, respectively. The dashed lines indicate our calculation with $N_{\rm test}=20$ test particles, where locality is restored approximately. Without test particles, we observe a loss of total angular momentum for each beam energy. 
	
\begin{figure}
		\centering
		\includegraphics[width=9.3cm]{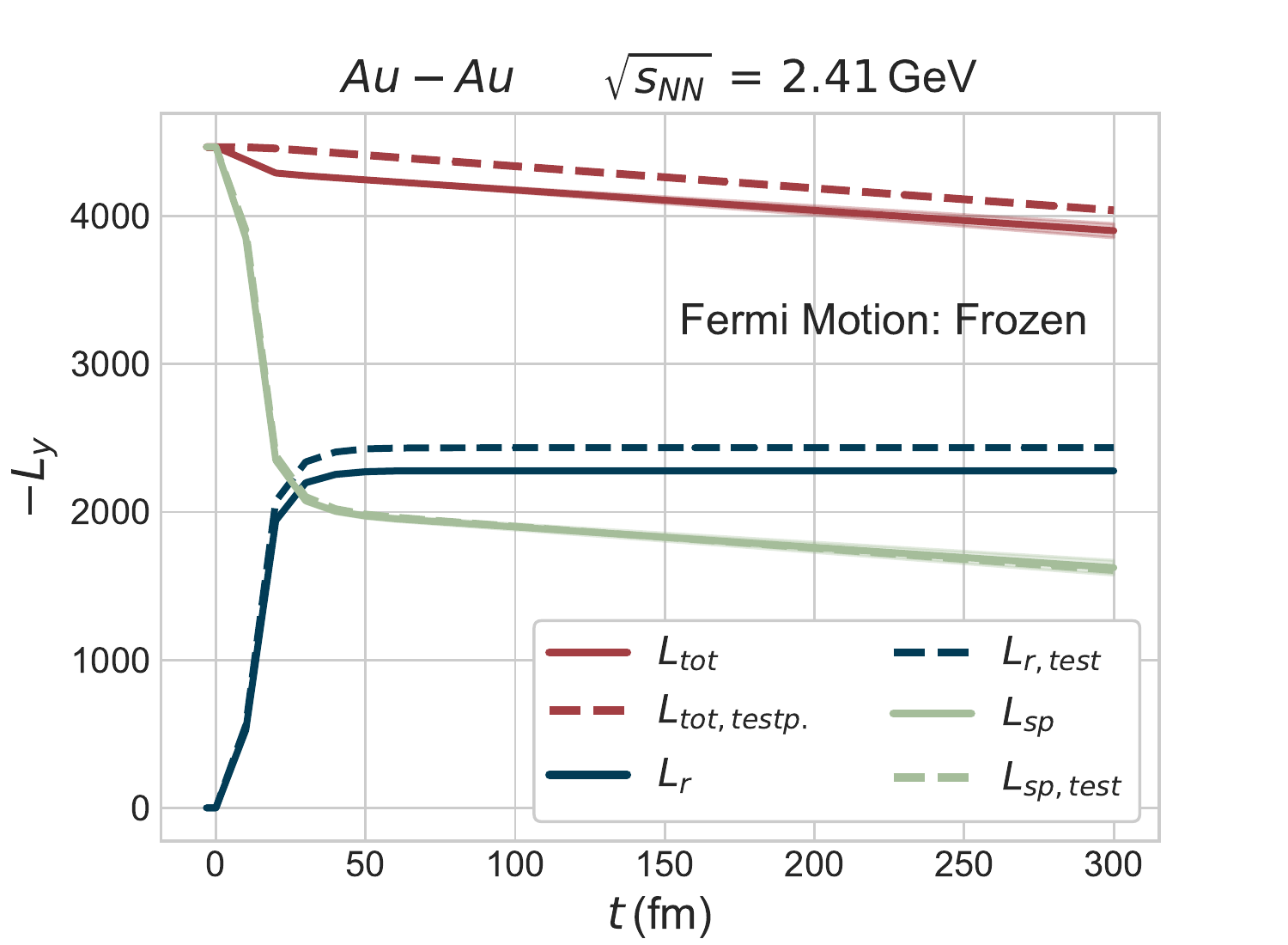}\vspace{14pt}
		\includegraphics[width=9.3cm]{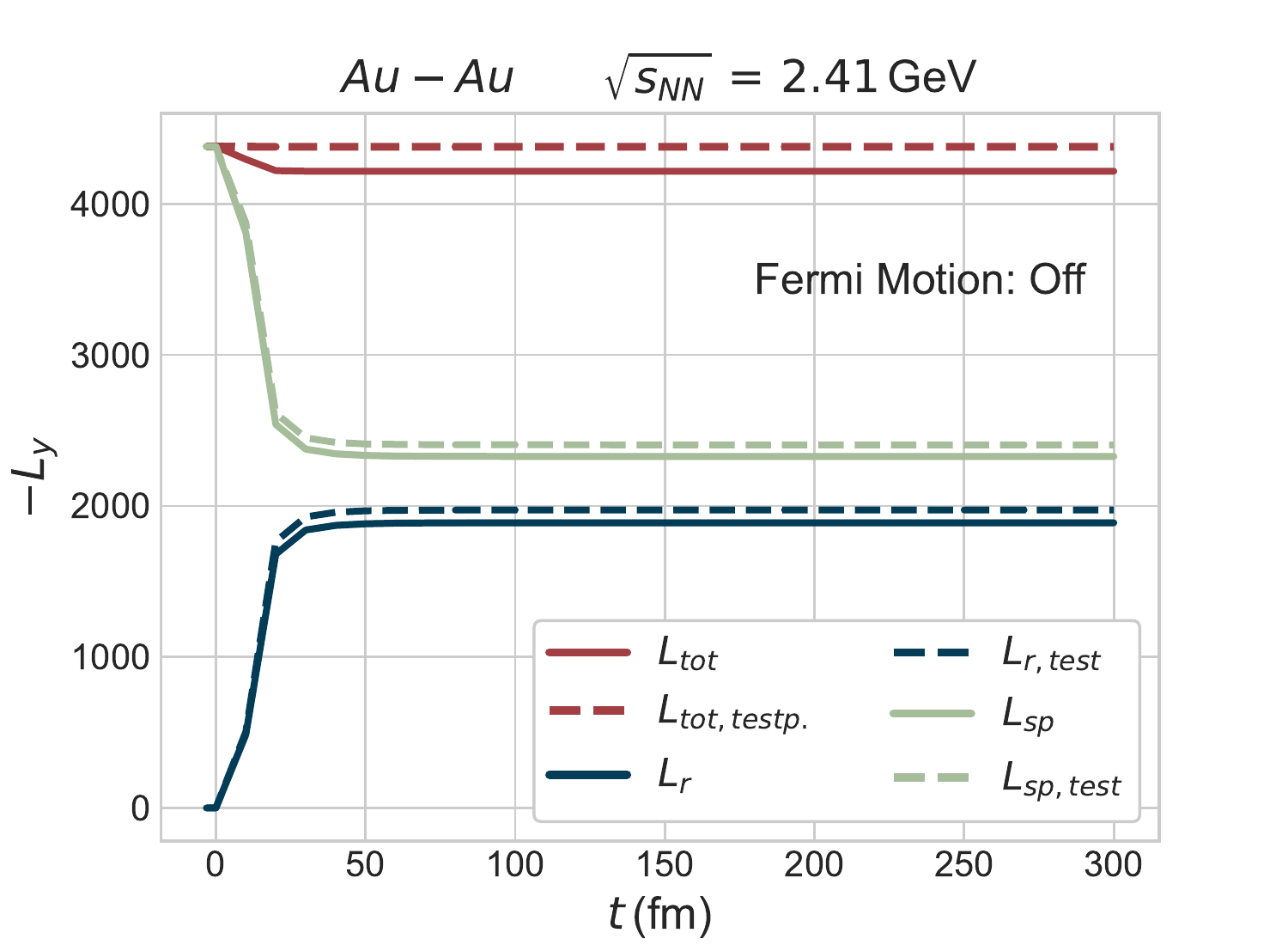}\vspace{14pt}
		\includegraphics[width=9.3cm]{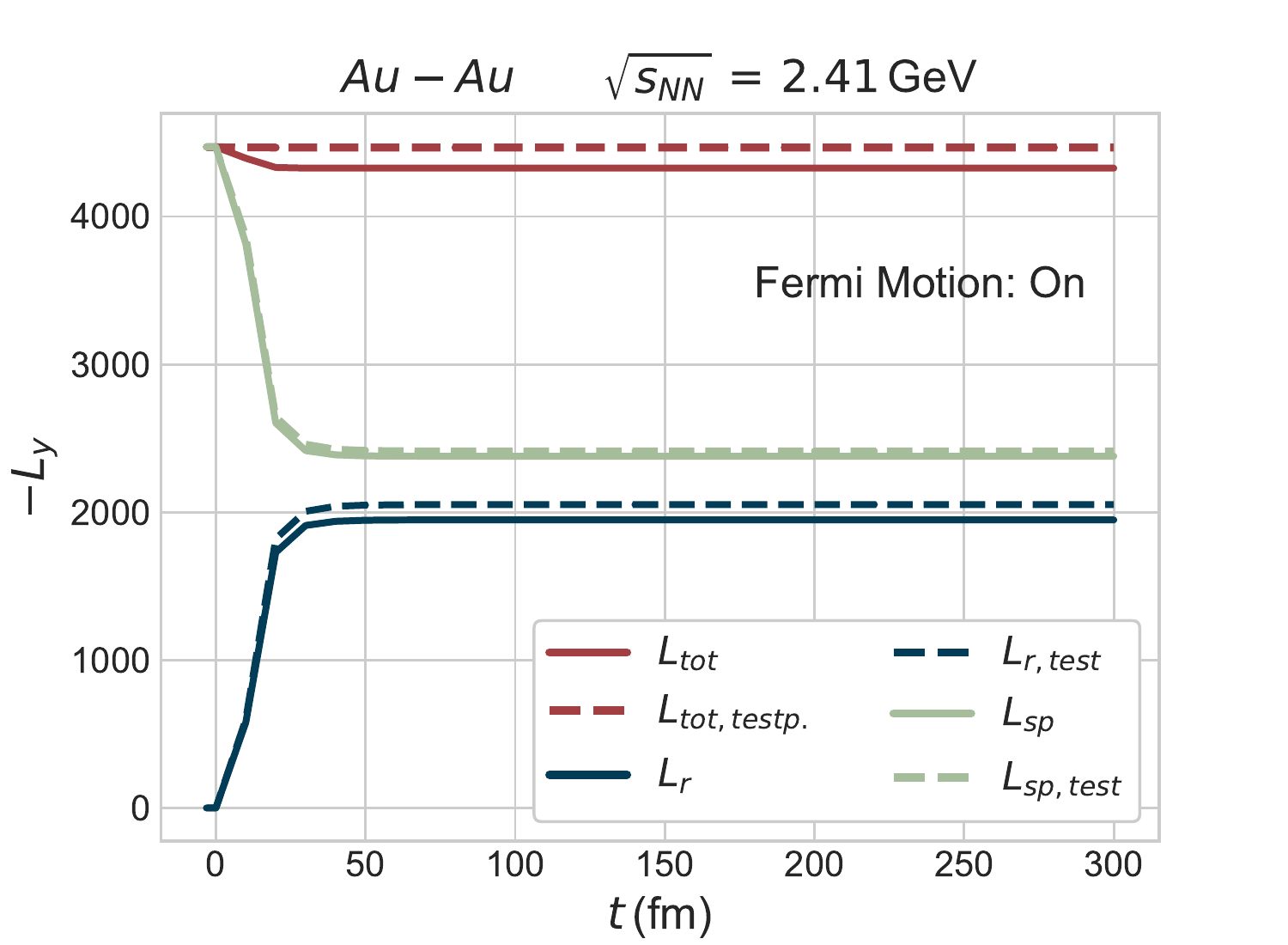}
		\caption{Time evolution of the angular momentum in AuAu collisions at $\sqrt{s_{\rm NN}}=2.41$ GeV. Compared are three different approaches for Fermi motion (frozen, off, on)
			The dashed lines depict the result for the same simulations including 100 test particles. It can be seen that the violation of angular momentum has a contribution originating solely from the frozen Fermi approach.}
		\label{fig:time_evolution_fermi_motion}
	\end{figure}

	The broken conservation does not have a single reason, but is an effect composed of two separate contributions: 
	as stated in Sec. \ref{sec:model}, our hadronic transport approach is based upon a geometrical interpretation of cross-sections as a maximum threshold for performing an interaction between incoming particles. Such an immediate finite range interaction breaks Poincaré invariance in binary interactions locally. Therefore, it also violates angular momentum conservation which can be seen from the kink in the total angular momentum and the subsequent undershooting during evolution \cite{Cheng_2002}. If test particles are added, the picture changes considerably. The time evolution of the angular momentum in Fig. \ref{fig:time_evolution} shows that the test particle method improves the non-conservation of angular momentum already for 20 test particles by correcting the participant contribution significantly, while the spectator's evolution stays nearly unchanged within error bars. This is particularly evident at early collision times, for which the initial kink is clearly smoothed out. We find that the non-conservation is purely driven by the participant evolution, which is consistent with our previous statement that the geometrical interpretation of cross-sections in binary interactions induces a spurious contribution to the angular momentum. 
	\begin{figure}[t]
		\centering
		\includegraphics[width=8.6cm]{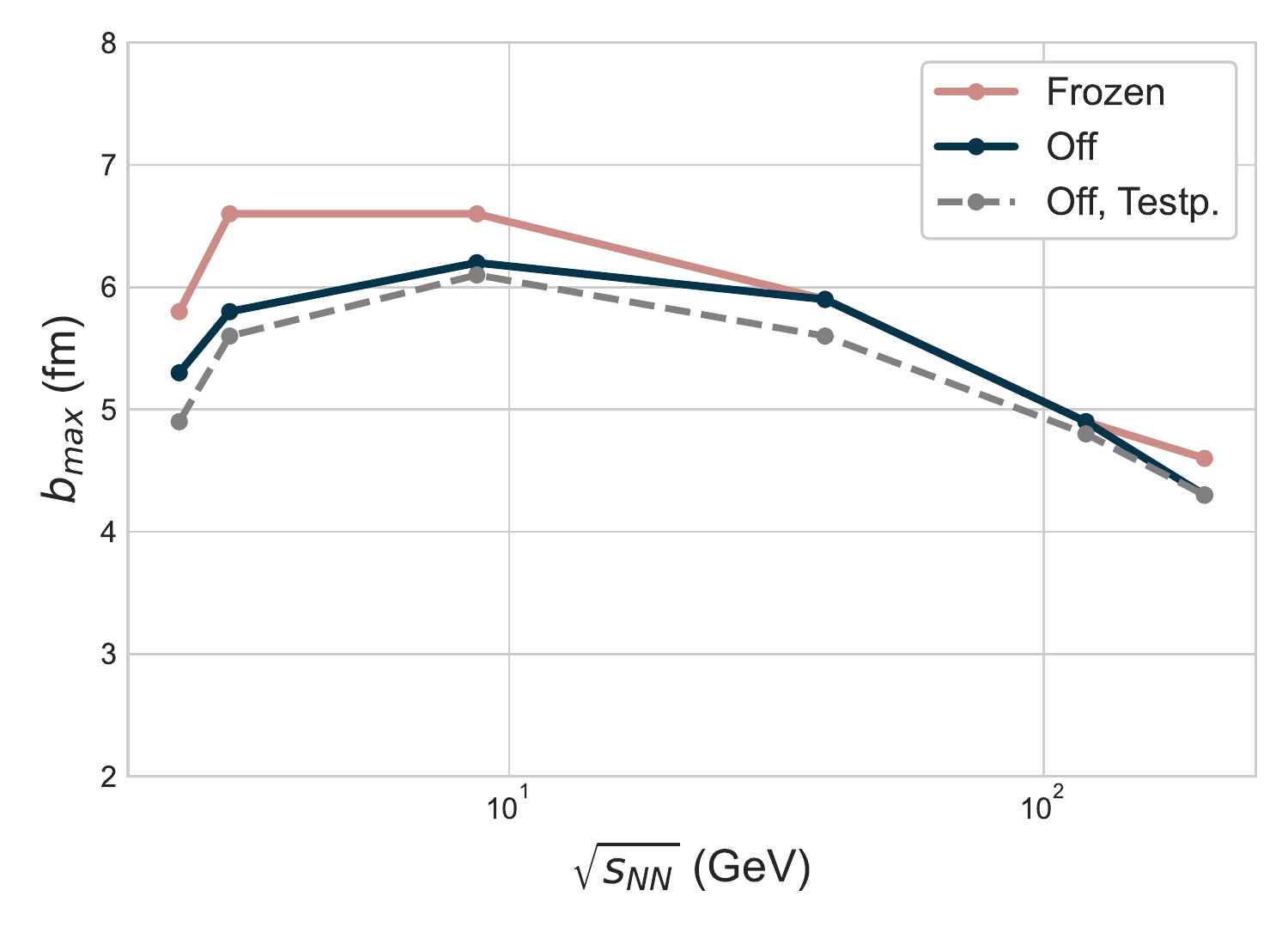}
		\caption{Impact parameter at which transfer of angular momentum is maximal as a function of the beam energy in AuAu collisions. Shown is the angular momentum of the participants with frozen Fermi motion and without Fermi motion. Additionally, the dashed line shows our results taking into account 20 test particles while Fermi motion is turned off.}
		\label{fig:energy_dep}
	\end{figure}
	However, especially for small beam energies as in the leftmost plot of Fig. \ref{fig:time_evolution}, we observe that the conservation of angular momentum can only be improved to a certain extent with the test particle method and a residual contribution exists that breaks the conservation, which does not appear to originate from the non-locality of binary interactions. This part is due to the frozen Fermi approach. With frozen Fermi motion, Fermi momenta are neglected for initial propagation of the nuclei and are considered only for interactions. This treatment leads to a discontinuity in the nucleon momenta, which consequently also contributes to the violation of angular momentum conservation. Accordingly, if this assertion is true, the angular momentum of the system must be conserved if, in addition to the test particle method, Fermi momenta are either neglected (off) or considered for the entire evolution, including initial propagation (on). Our statement is supported by Fig. \ref{fig:time_evolution_fermi_motion}. Here, similar to Fig. \ref{fig:time_evolution}, we show the angular momentum evolution of an AuAu collision at $\sqrt{s_{\rm NN}}=2.41\text{ GeV}$ with (dashed lines) and without test particles (solid lines). In each plot we applied Fermi motion in a different approach (frozen, off, on). As claimed earlier, we observe that conservation of angular momentum can be fully recovered with test particles if Fermi momenta are either neglected or considered throughout the whole evolution, starting from the initial state.
	In general, we find that the impact of broken conservation on the global behavior is not relevant for medium to high beam energies even without test particles. Nevertheless, as a conclusion, we can define optimal configurations for angular momentum studies in transport approaches. 
	In the low energy range, the test particle method in combination with Fermi motion taken into account over the entire evolution has proven to be the optimum in our studies. In this setup, however, potentials need to be taken into account to prevent the nuclei from flying apart.
	For medium to high beam energies, on the other hand, we see that the violation of angular momentum conservation predominately arises from the non-locality of the binary interactions and is thus remedied by the test particle method alone, while the contribution from the frozen Fermi approach is negligible.
	\subsection{Energy and system size dependence}
	\begin{figure}[t]
		\includegraphics[width=0.49\textwidth]{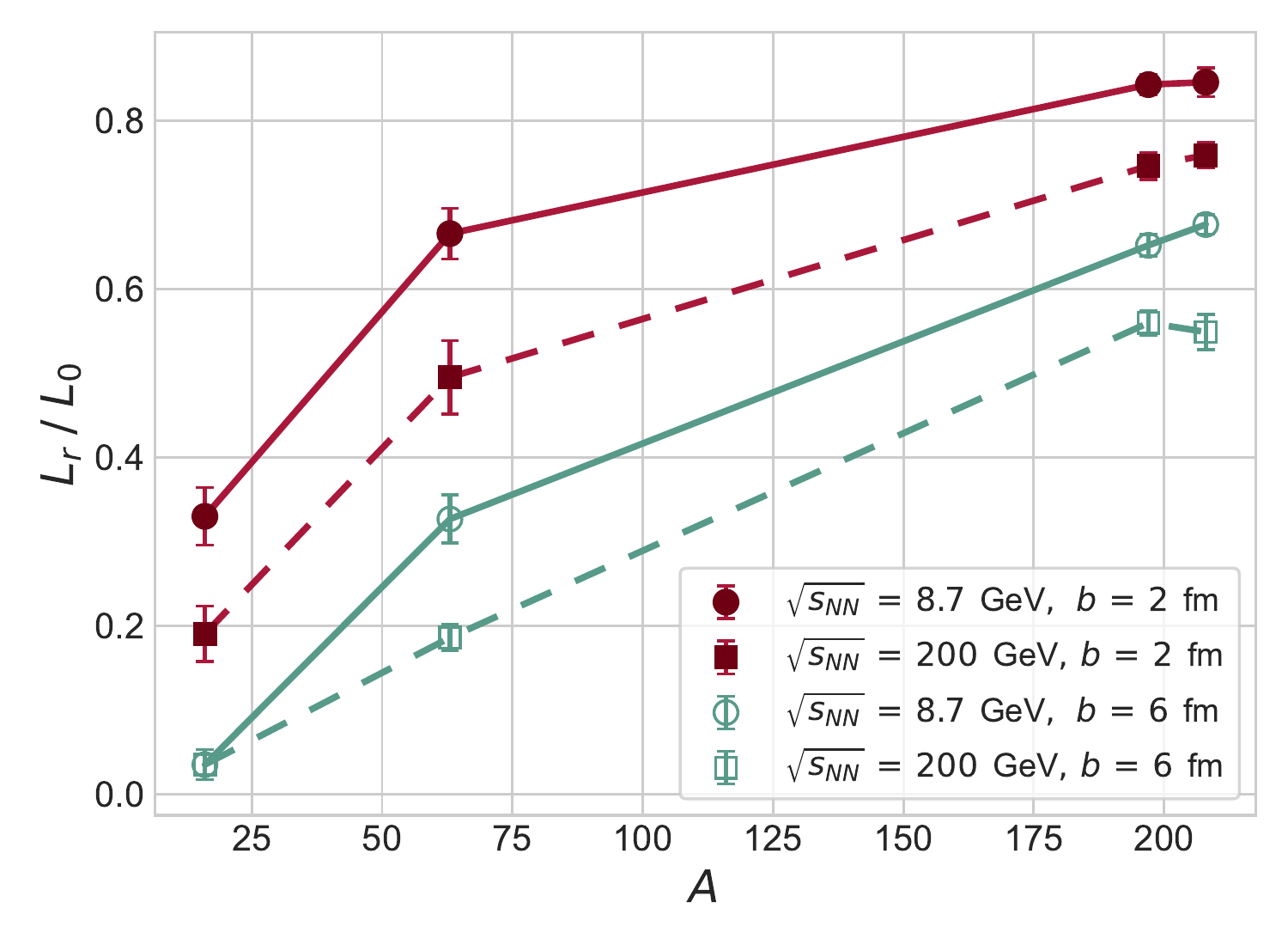}
		\caption{Ratio of the transferred angular momentum to the initial value as a function of system size and beam energy for two different fixed impact parameters.}
		\label{fig:system_size}
	\end{figure}
	From Fig. \ref{fig:imp_energy} we already concluded that we expect a single impact parameter that maximizes the angular momentum transfer at a given beam energy. However, we also want to understand whether such a maximum impact parameter is a generic characteristic that only depends on the structure of the initial state or if it changes as a function of the beam energy. 
	Therefore, we extract for each simulated beam energy, i.e. $\sqrt{s_{\rm NN}}=\left\lbrace 2.41, \,3.0, \,8.7, \,39.0, \,120.0,\, 200.0\right\rbrace \, \text{GeV}$, the corresponding $b_{max}$ and plot it as function of $\sqrt{s_{\rm NN}}$, as seen in Fig. \ref{fig:energy_dep}. The two solid curves depict our findings for the maximum impact parameter with Fermi motion (upper) and without (lower), respectively, while the dashed gray line represents our findings without Fermi motion and including 20 test particles. Our simulations show that the impact parameter of maximum angular momentum transfer depends weakly on the beam energy, precisely it varies within a range of $b_{max}\in \left[ 4.3\, \text{fm}, 6.2\, \text{fm}\right]$ without Fermi motion and $b_{max}\in \left[ 4.6\, \text{fm}, 6.6\, \text{fm}\right]$ including frozen Fermi motion.
	As already stated in Sec. \ref{subsec:results_impact_parameter}, also Fig. \ref{fig:energy_dep} reinforces our observation that Fermi motion shifts the point of maximum angular momentum transfer towards slightly more peripheral collisions.
	On the other hand, the curve for test particles consistently lies slightly below those without test particles, as we have also already seen in Sec. \ref{subsec:results_impact_parameter} due to the reduced cross-sections of the nucleons.\\
	Another quantity that determines the deposition of angular momentum in the fireball is the system size. In Fig. \ref{fig:system_size} we show our results for the system size dependence of the relative angular momentum transfer at mid-rapidity for different sets of fixed beam energies and impact parameters. To obtain Fig. \ref{fig:system_size}, we simulate a heavy-ion collision for nuclei with varying nucleon number in the range of $A = 16$ $(^{16}_8O)$ to $A = 208$ $(^{208}_{82}Pb)$. For each event we choose every combination between $\sqrt{s_{\rm NN}}={8.7\text{ GeV}, 200.0\text{ GeV}}$ and $b={2.0\text{ fm}, 6.0\text{ fm}}$. We calculate the relative transfer by normalizing the participant angular momentum to the angular momentum of the initial state.
	Thereby, the solid lines belong to our results for $\sqrt{s_{\rm NN}}=8.7\text{ GeV}$ and the dashed lines depict the case of $\sqrt{s_{\rm NN}}=200.0\text{ GeV}$.
	Although one might make the intuitive assumption that the deposition of angular momentum increases at higher beam energies due to the higher particle production and interaction rate, we see a clear trend towards lower beam energies and more central collisions for a higher relative angular momentum transfer. \\
	This result sets an important orientation marker to guide future experiments in the choice of systems and centralities, with the aim of identifying potential signals of a phase transition.
	
	\section{Conclusions} 
	\label{sec:conclusions}
	The evolution and transfer of angular momentum in heavy-ion collisions has been studied applying the hadronic transport approach SMASH. It was shown that the angular momentum of the fireball exhibits a distinct maximum for one unique impact parameter which is, furthermore, found to be a system specific parameter that is approximately beam energy independent. This is consistent with Glauber model predictions by Becattini \textit{et al} \cite{Becattini_2008}. System size studies at mid-rapidity suggest that the relative transfer and deposition of angular momentum in the interaction volume is larger for lower beam energies and in more central collisions. This observation is made equally for all studied nucleus sizes.
	Furthermore, our results show that the impact of the non-conservation of angular momentum due to Poincaré violations originating from the geometrical interpretation of cross-sections and the frozen Fermi approach are insignificant, especially in the medium to high beam energy regime. 
	However, in order to restore locality fully, we applied the test particle method as a simple tool to guaranty angular momentum conservation in the limit $N_\text{test}\rightarrow\infty$. Even though finite Fermi momenta increase the angular momentum at all beam energies, we found that they are most relevant for low beam energies as expected. 
	Our results indicate under which conditions the highest transfer of angular momentum in a heavy-ion collisions is expected, which is relevant to guide experimental programs and help to identify possible signals of a phase-transition of hadronic matter by constraining the dynamical evolution. In the future, the determination of vorticity based on coarse-graining the transport evolution could be interesting to compare to experimental data for $\Lambda$ polarization. 
	
	\begin{acknowledgments}
		This work was funded by the Deutsche Forschungsgemeinschaft (DFG, German Research Foundation) – Project number 315477589 – TRR 211. Computational resources have been provided by the GreenCube at GSI. H.E. and N. S. acknowledge the support by the State of Hesse within the Research Cluster ELEMENTS (Project ID 500/10.006).
	\end{acknowledgments}
	
	\pagebreak
	
	\bibliography{AngularMomentum_arXiv.bib}
	
\end{document}